**Probing spin order via magnon transmission across quantum Hall ferromagnet heterojunctions**

Seung Hwan Lee[1,a], Shaowen Chen[1,b], Andrew T. Pierce[1,c], Patrick R. Forrester[1], Kenji Watanabe[2], Takashi Taniguchi[3], Amir Yacoby[1‡]

[1]*Department of Physics, Harvard University, Cambridge, MA 02138, USA*

[2]*Research Center for Electronic and Optical Materials, National Institute for Materials Science, 1-1 Namiki, Tsukuba 305-0044, Japan*

[3]*Research Center for Materials Nanoarchitectonics, National Institute for Materials Science, 1-1 Namiki, Tsukuba 305-0044, Japan*

[a] *Present address: Department of Physics, Yale University, New Haven, CT 06520, USA*
[b] *Present address: Department of Physics and Astronomy, Rutgers University, Piscataway, New Jersey 08854, USA*
[c] *Present address: Kavli Institute at Cornell for Nanoscale Science, Ithaca, NY, USA*
[‡]Corresponding author's email: yacoby@g.harvard.edu

**Two-dimensional material platforms now host a remarkable array of exotic correlated phases, from unconventional superconductivity[1–6] to fractional Chern insulators[7,8]. Probing magnetic order in these systems is essential for understanding their underlying physics, yet dilute spin densities render conventional magnetic probes ineffective. Spin waves, or magnons, in quantum Hall ferromagnets (QHFM)[9–12] have proven effective for probing the magnetic order in various symmetry-broken quantum Hall (QH) phases in graphene systems[13–19], but previous works have been limited to homojunction configurations within a single material. Here, we demonstrate magnon transmission across a monolayer-bilayer graphene quantum Hall ferromagnet heterojunction—the first magnon transmission across quantum Hall ferromagnet heterojunctions, using one material as a magnon source to probe magnetic order in a distinct material. Generating magnons in monolayer graphene (MLG) at ν = 1, we detect their transmission through bilayer graphene (BLG) via nonlocal voltage measurements, revealing spin order in BLG symmetry-broken quantum Hall states. The transmission exhibits hallmark magnon signatures: a sharp onset at the Zeeman energy and systematic variation with Landau level filling, including suppression at ν = 4 and 8 where spin polarization vanishes. Our findings establish heterojunction magnon transmission as a**

**powerful, modular probe of magnetic order, opening new avenues for investigating exotic quantum states across the rapidly expanding family of two-dimensional materials.**

Quantum Hall ferromagnetism in graphene systems[10,20] provides an ideal platform for studying symmetry-broken electronic phases in two dimensions. Recent advances in probing spin order through magnon transport[13–17] in monolayer graphene (MLG) quantum Hall ferromagnets (QHFMs) have revealed rich physics of symmetry-broken states. This success highlights the potential for magnon-based probes to investigate magnetic order in the rapidly expanding family of emerging quantum materials—including twisted bilayer graphene and rhombohedral graphene—that host exotic phases including unconventional superconductivity[1–6], quantum anomalous Hall states[5,21], and fractional Chern insulators[7,8]. These correlated phases have attracted intense scientific interest due to their fundamental importance and technological applications. However, the extremely dilute spin density inherent to these systems makes conventional magnetic probes—such as X-ray scattering, neutron scattering, and ferromagnetic resonance—difficult to apply, calling for new experimental approaches to reveal their magnetic order. Extending magnon-based techniques across heterojunctions—using one material as a magnon source to probe another—would significantly broaden the range of systems accessible to this approach.

Here, we demonstrate magnon transmission across an MLG/BLG quantum Hall ferromagnet heterojunction—the first realization of electrically-generated spin waves probing magnetic order in a dissimilar two-dimensional material. By launching magnons from a $\nu_{MLG} = 1$ MLG QHFM source and detecting their transmission through BLG quantum Hall states, we establish a novel approach to reveal spin order in symmetry-broken phases in BLG. The nonlocal transport signal measured within the BLG region exhibits hallmark signatures of magnon transport: it appears only in Landau level fillings where the spin-valley symmetry is broken and exhibits a sharp onset at the Zeeman energy. This technique establishes magnon transmission as a powerful new probe of magnetic order across the expanding landscape of two-dimensional quantum materials.

**Experimental setup**

To probe spin wave transmission across MLG/BLG QHFM, we studied two graphene devices (Device A and B) which contain both MLG and BLG regions. In the main text, we focus on device A. The measurement scheme is described in Fig 1. The bulk filling factor of MLG is tuned to $\nu_{MLG} = 1$, creating a ferromagnetic bulk for magnon generation at the MLG contacts region[13–16,18,20]. Magnons are generated in the MLG region by applying a DC bias voltage $|V_{DC}| > E_Z/e$ to the source contact and freely propagate to the MLG/BLG junction via the ferromagnetic MLG bulk. As in previous studies[13,14], magnon generation is accompanied by a sharp deviation in the two-terminal conductance $g_{MLG}$ from the quantized value of $e^2/h$. In our devices, some of the magnons enter the BLG bulk region to be ultimately absorbed at the BLG contacts, inducing a nonlocal voltage in the BLG region which serves as a signature of magnon transmission. To verify the magnon transmission, we monitor the two-terminal conductance $g_{MLG}$ between the generation contacts within the MLG as well as the nonlocal voltage between the detection contacts within BLG as the DC bias $V_{DC}$ across the magnon generation contacts is varied. We note that both electrical generation and detection of magnons in QHFM require two oppositely spin-polarized edge states at magnon generation/absorption sites. In generation sites (MLG contacts), this condition is conveniently satisfied due to contact doping[13]. In the detection sites (BLG contacts), we maintain filling factor $\nu_{BC} > 4$ to ensure the presence of oppositely spin-polarized edge states required for magnon absorption using the silicon back gate. Due to the observed *n*-doping of the contacts, we restrict our analysis to *n*-doped carrier densities of both MLG and BLG.

To minimize disorder at the MLG/BLG junction, we choose a flake with straight, naturally exfoliated boundary between MLG and BLG regions (Fig 1c inset). Patterned top and bottom metallic gates and a *p*-doped Si global back gate (SiG) control the carrier densities in the MLG bulk, BLG bulk, and BLG contact regions independently (see Methods). During device fabrication, extra steps were taken (see Methods and Extended Fig. 1) to carefully align the TG edge to the MLG/BLG boundary, and we estimate the alignment error to be within the nominal lithographic tolerance of 50 nm. The BLG region exhibit symmetry-broken integer quantum Hall incompressible states for $0 < \nu_{BLG} < 24$ and fractional quantum Hall (FQH) states for $0 < \nu_{BLG} < 8$ (Fig 1e), demonstrating the high quality of the device. All presented measurements are performed at the base temperature of at 20 mK in a dilution refrigerator.

**Magnon transmission across MLG/BLG QHFM heterojunctions**

We demonstrate magnon transmission into BLG via two complementary signals (Fig. 2): the two-terminal conductance $g_{MLG}$ in the MLG region and the nonlocal voltage signal $S_{NL} =$ $dV_{NL}/dV$ in the BLG region. Here $V = V_{DC} + V_{AC}$ is the voltage applied between the magnon generation sites in MLG, where $V_{AC}$ is a small AC excitation. As $V_{DC}$ is varied at $B = 9$ T, and with $\nu_{MLG} = 1$ fixed into the ferromagnetic incompressible state, $g_{MLG}$ exhibits a sharp deviation from $e^2/h$ at the Zeeman energy $V_{DC} = E_Z/e = g\mu_B B/e \approx 1\ mV$, heralding the onset of magnon generation in the MLG region. Crucially, this signature is found to be independent of the carrier density in the BLG region (Fig. 2b), consistent with the expectation that $g_{MLG}$ is mainly determined by magnon generation and absorption processes taking place only in the MLG region (see Methods).

In order to examine the effect of the impinging magnons on the BLG bulk, we monitor the nonlocal signal $S_{NL}$ as a function of $V_{DC}$ as $\nu_{BLG}$ is varied. Remarkably, $S_{NL}$ becomes nonzero at the Zeeman threshold for magnon generation in MLG (Fig. 2d), demonstrating that magnons generated in the MLG sector successfully propagate across the heterojunction and through the BLG bulk to be detected at the BLG contacts. This interpretation is further corroborated by measurements of $S_{NL}$ at different magnetic fields (Figs. 2f-g), which show that the onset of $S_{NL}$ closely follows $E_Z$ and that the signal is strongly suppressed whenever $|V_{DC}| < E_Z$. We emphasize that, because $g_{MLG}$ is independent of $\nu_{BLG}$ and shows no correlation with $S_{NL}$ when $\nu_{BLG}$ is varied, any direct leakage current contributions to $S_{NL}$ must be minimal; moreover, $S_{NL}$ exhibits no clear correlation with longitudinal resistance $R_{xx}$ measured at the same contact pair (Fig. 2e), ruling out the possibility that $S_{NL}$ merely reflects charge transport between these contacts. We thus conclude that magnon transmission from the MLG QHFM into the BLG bulk is the origin of the nonlocal signal.

The correlation between magnon generation and transmission is further established by varying a complementary set of parameters—$\nu_{MLG}$ between $\nu_{MLG} = 0$ and 2 under different DC bias conditions (Fig. 3). In the absence of DC bias ($V_{DC} = 0$ mV), the nonlocal signal does not appear (Fig. 3b) as magnons are not generated in MLG. In contrast, we observe a robust nonlocal signal by applying $V_{DC} = -3$ mV (Fig. 3c), with the signal most pronounced within the density range corresponding to the $\nu_{MLG} = 1$ conductance plateau (between the two dashed lines). The signal extends to MLG densities below the $\nu_{MLG} = 1$ plateau where the two-terminal conductance remains finite, though the signal magnitude is reduced compared to within the plateau. This persistence suggests that magnon generation can occur outside the incompressible state. Such persistence requires both edge and bulk states to retain partial spin polarization necessary for magnon generation and propagation, suggesting that ferromagnetic order persists in the compressible regime. On the other hand, the signal becomes strongly suppressed as MLG density approaches the $\nu_{MLG} = 2$ conductance plateau even when $|V_{DC}| > E_Z$, consistent with the suppression of both magnon generation and transmission due to the absence of spin order at $\nu_{MLG} = 2$. One interesting feature is that the nonlocal signal is strongly suppressed as soon as $g_{MLG}$ increases beyond the $\nu_{MLG} = 1$ plateau even if it has not reached $\nu_{MLG} = 2$ plateau. This suppression arises as both the bulk spin polarization and generation efficiency decrease in the transition toward $\nu_{MLG} = 2$. Previous experiments[13,15,16] also reported similar phenomenology in which magnon transmission across the MLG homojunction with fillings $1 < \nu_{MLG} < 2$ was strongly suppressed compared to $0 < \nu_{MLG} < 1$, further reinforcing that the nonlocal signal arises from magnon transmission across the MLG/BLG heterojunction.

**BLG Landau level filling dependence in magnetic field**

Having demonstrated magnon transmission across the MLG/BLG QHFM heterojunction, we now show that the magnon transmission characteristics directly probe the magnetic order in BLG QHFM states. Figure 4a, b present $S_{NL}$ as a function of $\nu_{BLG}$ and perpendicular magnetic field $B$ ranging from 4-9 T. While the magnetic field is swept, MLG density is maintained within the $\nu_{MLG} = 1$ conductance plateau to ensure robust magnon generation (Extended Data Fig. 7).

Consistently over the entire field range of 4 T to 9 T, we observe signatures of magnon transmission at symmetry-broken BLG LL fillings. The nonlocal signals $S_{NL}$ are pronounced under a finite voltage bias (Fig. 4a) but vanish when the bias is turned off and no magnons are generated (Fig. 4b). $S_{NL}$ exhibits numerous prominent vertical features (Fig. 4a), corresponding to magnon absorption features that occur at fixed Landau level filling, consistent with the expectation that $S_{NL}$ probes the properties of the various BLG bulk quantum Hall phases that are stabilized for $0 < \nu_{BLG} < 8$. The polarity of $S_{NL}$ reflects the differential magnon absorption between two contact sites, with the sign indicating preferential absorption at one site versus the other[13] (see Methods). An intriguing feature of $S_{NL}$ is the systematic polarity reversal observed as $\nu_{BLG}$ crosses integer values: for most integer fillings in the range $0 < \nu_{BLG} < 8$, we observe positive $S_{NL}$ at $n - \varepsilon$ and negative $S_{NL}$ at $n + \varepsilon$ for small $\varepsilon > 0$. This polarity pattern cannot be explained solely by geometric considerations of absorption site proximity[13] because the device geometry remains fixed while $\nu_{BLG}$ is varied. Polarity reversals in magnon absorption signals have been observed in MLG experiments as a function of DC bias[16], suggesting that magnon absorption processes can be sensitive to non-geometric origin. We speculate that filling-dependent changes in the local electronic environment may modulate the relative absorption efficiency between the two absorption sites, though the precise mechanism remains an open question. Additionally, while finite signals extend down to $\nu_{BLG} \sim 0$ where complex magnetic phases are predicted, the transport signatures at these lower fillings show anomalous features that distinguish them from the magnon transmission behavior observed at higher fillings and in previous works[13–16,18]. Our analysis therefore focuses on $\nu_{BLG} \geq 0.7$ where signatures are consistent with established magnon physics.

Close examination of $S_{NL}$ at selected values of magnetic field (Fig. 4c) shows, however, that the nonlocal signal abruptly vanishes near fillings $\nu_{BLG} = 4$ and 8, despite being nonzero at all other fillings in the range of interest. The vanishing of $S_{NL}$ at $\nu_{BLG} = 4$ and 8 has been reproduced under a wide range of conditions, including at different bias voltages (Fig. 2 and Extended Data Fig. 3), contact configurations (Extended Data Fig. 5-6), and in a second device (Extended Data Fig. 9-11), calling for an explanation of the effect based on the intrinsic properties of the BLG ground states at $\nu_{BLG} = 4$ and 8.

We first consider the behavior of $S_{NL}$ at partial fillings in the lowest LL ($0 < \nu_{BLG} < 4$). Our measurements reveal finite $S_{NL}$ between each integer filling, with pronounced enhancement near integer values. This behavior aligns with predictions of the spin order and previous experimental studies specifically in the case of a small displacement field[22–27] relevant for our measurements (see Methods and Extended Data Fig.4), according to which the negative (positive) fillings of the BLG lowest LL are occupied by spin ↑ (↓) electrons. Within this picture (Fig. 4d), a ground state with broken spin symmetry—a prerequisite for magnon transmission—is expected for all $0 < \nu_{BLG} < 4$. However, once $\nu_{BLG} = 4$ is reached, the eight filled levels of the $N = 0, 1$ LL manifold together form a spin-valley-orbital singlet, thus resulting in a nonmagnetic ground state through which magnons cannot propagate, thus eliminating the nonlocal signal. While consistent with this framework, our observations cannot definitively exclude alternative spin configurations arising from purely interaction-driven models[28], in which the valley splitting is the largest energy scale and spins are filled in the order ↑, ↑, ↓, ↓ ($\nu_{BLG}$ =1, 2, 3, 4). Regardless of the specific level ordering our measurements establish unambiguously that spin symmetry remains broken for all fillings between $\nu_{BLG} = 0$ and $\nu_{BLG} = 4$.

For the $N = 2$ LL ($4 < \nu_{BLG} < 8$), nonlocal signals emerge near integer fillings $\nu_{BLG} = 5, 6, 7$. In particular, the presence of a nonzero $S_{NL}$ at $\nu_{BLG} = 6$ is consistent only with filling taking place in the sequence $K'\uparrow, K\uparrow, K'\downarrow, K\downarrow$ predicted to be appropriate for the small displacement field regime[22–27]. Notably, the $N = 2$ LL signals exhibit qualitatively different behavior compared to the $N = 0$ and 1 LLs: the nonlocal signal occurs predominantly near the integer values, with significant suppression between integers (i.e., at non-integer Landau level fillings). This qualitative difference may be closely related to differences in ground state order, as demonstrated in MLG[16] where underlying ground state phases can dramatically alter magnon transmission at non-integer fillings. While the additional orbital degeneracy of the $N = 0$ and $N = 1$ levels may further contribute to this difference, the origin of the contrasting transmission characteristics remains an open question that could illuminate the interplay between ground state physics and magnon propagation. These systematic measurements establish magnon transmission spectroscopy as a powerful probe of magnetic order in quantum Hall systems, providing access to spin polarization in ground states that are challenging to examine with conventional techniques.

**Discussion and Outlook**

We have demonstrated that magnon transmission using MLG QHFM as a spin wave source provides direct access to magnetic order in symmetry-broken BLG quantum Hall states. In this all-electrical architecture, the MLG $\nu = 1$ ferromagnet serves as the magnon source, while detection at the BLG contacts reveals the filling-dependent spin polarization of the BLG. The observed transmission characteristics—the sharp onset of transmission at the Zeeman energy and its systematic dependence on Landau level filling—establish magnon transmission as a powerful new technique for probing magnetic order. Additionally, the displacement field dependence of magnon transmission (Supplementary Fig. 2) suggests potential for probing layer polarization in BLG, though further investigation is needed to establish this connection definitively.

Our approach is fully electrical and gate-defined, requiring only oppositely spin-polarized edge states for magnon generation and detection—conditions naturally satisfied by many quantum Hall-like systems. This modular architecture, where a well-characterized magnon source probes a distinct target material, extends the reach of magnon-based techniques to systems that cannot efficiently generate magnons on their own. The rapidly expanding family of moiré materials and rhombohedral graphene systems represents promising platforms for such studies, as these gate-tunable systems can host both zero-field quantum Hall-like phases and exotic correlated phases such as fractional Chern insulators[8,29–31], offering the potential for magnon transmission spectroscopy to reveal spin order across this diverse range of emergent quantum states. Recent advances in in-situ control of moiré twist angle[32,33] present particularly exciting opportunities to naturally integrate MLG with moiré systems, enabling systematic investigation of magnetic order as a function of twist angle through magnon transmission measurements to probe spin order in emergent phases[34]. Our demonstration of heterojunction magnon transmission opens new pathways for investigating exotic quantum states across diverse two-dimensional platforms beyond homojunction configurations.

## Methods

**Sample preparation.** We fabricated MLG/BLG heterostructure devices by carefully selecting graphene flakes containing both monolayer and bilayer regions separated by clean, straight boundaries. The devices were assembled using a dry transfer technique[35], where the graphene was encapsulated between two hexagonal boron nitride (hBN) crystals (top: 35 nm, bottom: 70 nm) on a pre-patterned PdAu local back gate. The 20 nm thick PdAu gate was thermally evaporated and vacuum annealed at 400°C for 8 hours to ensure pristine surface quality. Critical to device operation was precise alignment of the top gate to the MLG/BLG boundary, which we achieved through careful registration using additional alignment marks evaporated near the device area after encapsulation. The top hBN surface was cleaned using an AFM tip before depositing a 55 nm gold top gate over the BLG bulk region. Electrical contacts were made using angle-evaporated Cr/Au, with selective etching of only the top hBN layer. This selective etching was crucial to prevent shorting for MLG contacts positioned above the PdAu gate. The final device geometry was defined by reactive ion etching with $CHF_3/O_2$ plasma using a PMMA mask.

Our design takes advantage of contact doping to establish $\nu_{MLG} = 2$ edge states near the MLG contacts, while BLG contacts were deliberately placed outside the PdAu gated region. We note that devices with irregular MLG/BLG boundaries exhibited inferior transport characteristics (e.g. only a limited number of symmetry broken states in $0 < \nu_{BLG} < 8$ were observed), possibly due to inhomogeneous strain fields created during mechanical exfoliation.

MLG bulk density is controlled by a PdAu metal backgate (BG); BLG bulk density is controlled by both BG and gold local top gate (TG); and BLG contact region is controlled by silicon global back gate (SiG).

**Transport Measurements.** Transport measurements were performed in a Leiden dry dilution refrigerator at a nominal base temperature of 20 mK. We employed standard low-frequency lock-in techniques using SR830 amplifiers and SR560 voltage preamplifiers for enhanced signal detection. All measurements used AC excitation voltages between 50-200 μV at 17.777 Hz.

**Calculation of density and displacement field.** The carrier density and displacement field were determined by first extracting gate capacitances from Landau fan measurements. The carrier density is calculated as

$$n = (c_b(V_b - V_{b0}) + c_t(V_t - V_{t0}))/e$$

where $c_b$ and $c_t$ are the back gate and top gate capacitances per unit area, $V_b$ and $V_t$ are the applied gate voltages, $V_{b0}$ and $V_{t0}$ are the gate voltages corresponding to charge neutrality, and $e$ is electron charge. The displacement field is calculated by

$$D = (c_b(V_b - V_{b0}) - c_t(V_t - V_{t0}))/2\epsilon_0$$

where $\epsilon_0$ is the vacuum permittivity.

In our device geometry, the bottom gate controls carrier density in both MLG and BLG regions, while the top gate only affects the BLG region. For most measurements, the MLG density was maintained at $\nu_{MLG} = 1$ to ensure robust magnon generation, which adds a constraint to the gate voltage combinations. Under this constraint, the BLG density and displacement field are controlled by varying the top gate voltage while adjusting the bottom gate to maintain the MLG filling factor. The range of displacement field values in our field sweep measurements is shown in Extended Data Fig. 4.

**Contacts resistance in calculating $E_Z$:** Series resistance including contact resistance must be accounted for when determining the Zeeman energy from transport measurements. We calibrated the series resistance using the quantized conductance plateaus at integer filling factors. The contact-corrected conductance is calculated as $g_{MLG} = R_Q/(\frac{V_{ac}}{I_{ac}} - R_c)$, where $V_{ac}$ is the applied voltage, $I_{ac}$ is measured current, $R_c$ represents the contact resistance and $R_Q = h/e^2$. The value of $R_c$ is determined by requiring that the measured conductance plateaus match the expected quantized values ($ne^2/h$, $n$ integer) at integer fillings. This contact resistance correction ensures accurate determination of the bias voltage at which magnon generation occurs, allowing precise identification of the Zeeman energy threshold $E_Z = g\mu_B B$.

**Nonlocal signals upon magnon absorption:** Following the framework established by Ref. [13], using current conservation and assuming perfect equilibration between edge modes and contacts, the measured differential voltage at contacts B2 and B3 can be expressed as:

$$S_{NL} = d\varepsilon_3/d\mu - d\varepsilon_2/d\mu$$

where $\varepsilon_i$ represents the change in edge channel chemical potential at absorption site i as shown in Fig. 1. The nonlocal signal thus measures the steady-state imbalance in chemical potential redistribution between the two absorption sites, which arises from preferential magnon absorption that depends on distance, geometry, and local electronic environment near each absorption site.

**Valley ordering in bilayer graphene Landau levels.** In bilayer graphene, the low-energy electronic states form Landau levels for which the single-particle energies are given by[22,36]:

$$E_0 = \frac{1}{2}\xi U + E_s\sigma$$

$$E_1 = \left(\frac{1}{2} - \frac{\hbar w_c}{\gamma_1}\right)\xi U + E_s\sigma + \Delta_{10}$$

$$E_{\pm N} = \pm\hbar w_c\sqrt{N(N-1)} - \frac{\hbar w_c}{2\gamma_1}\xi U + E_s\sigma \qquad (N \geq 2)$$

where $\sigma$ is spin, $\xi$ is the valley index (+1 for K, -1 for K'), $U$ is the interlayer potential difference $U = Dd_0$ ($d_0$: the interlayer distance, $D$: displacement field), $E_s$ represents Zeeman energy, and $\Delta_{10}$ is the orbital energy. From Slonczewski-Weiss-McClure band parameters[37], $\gamma_1 \sim 0.39$ eV, and thus $\hbar w_c/\gamma_1 < 1/2$ for magnetic fields $B < 9$ T used in our experiments. Therefore, the coefficients of the displacement field term $\xi U$ have opposite signs for the zero-energy Landau levels ($N = 0,1$) compared to higher Landau levels ($N \geq 2$). Specifically, $E_0$ and $E_1$ have positive coefficients (1/2 and 1/2 - $\hbar w_c/\gamma_1 > 0$), while $E_{\pm N}$ has a negative coefficient ($-\hbar w_c/2\gamma_1 < 0$). This opposite displacement field response is incorporated in our analysis between the $N$ = 0,1 and $N$ = 2 manifolds where the sign change in displacement field from positive to negative around $\nu_{BLG} = 4$ (Extended Data Figure 4).

**Note on accidental graphene/hBN alignment in device A.** Device A exhibits an accidental graphene/hBN superlattice with an estimated angle of 0.6 degrees, as determined from Brown-Zak oscillations in transport measurements (Supplementary Figure 1). This accidental alignment does not influence our observations and conclusions for several reasons. First, our measurements are conducted at magnetic fields and carrier densities well below the regime where Hofstadter butterfly physics becomes relevant. For the observed moiré pattern, the characteristic magnetic field scale for Hofstadter effects is $B \approx 34$ T (corresponding to one magnetic flux quantum per moiré unit cell), and the superlattice density is $n_s \approx 3.3 \times 10^{12}$ cm$^{-2}$. Our experiments are performed at significantly lower fields and densities, and the transport data shows no signatures of superlattice effects in $N = 0,1$ LL within this field range. Second, device B, which shows no evidence of graphene/hBN alignment, demonstrates magnon transmission signals in $N = 0,1$ LL that are consistent with those from device A, confirming that graphene/hBN crystallographic alignment does not affect our key observations.

**Data availability**

The data that supports the findings of this study are available from the corresponding authors upon reasonable request.

**Acknowledgements**

We are particularly grateful to Bertrand I. Halperin for insightful discussions, and also thank Yuan Cao and Nemin Wei for helpful discussions. A.Y. is supported by the Gordon and Betty Moore Foundation through Grant No. GBMF 12762, and by the U.S. Army Research Office (ARO) MURI project under Grant No. W911NF-21-2-0147. This work was performed, in part, at the Center for Nanoscale Systems (CNS), a member of the National Nanotechnology Infrastructure Network, which is supported by the NSF under award no. ECS-0335765. CNS is part of Harvard University. A.T.P. acknowledges funding from the Kavli Institute at Cornell for Nanoscale Science (KIC) Postdoctoral Fellowship. P.R.F. acknowledges support from the National Science Foundation Graduate Research Fellowship under grant number DGE 1745303. K.W. and T.T. acknowledge support from the JSPS KAKENHI (Grant Numbers 21H05233 and 23H02052), the CREST (JPMJCR24A5), JST and World Premier International Research Center Initiative (WPI), MEXT, Japan.

**Author Contributions**

S.H.L. and A.Y. designed the experiment. S.H.L. fabricated the device and performed the transport measurements in the dilution refrigerator and analyzed the data with input from A.Y. K.W. and T.T. provided hBN crystals. All authors participated in discussions and in writing the manuscript.

**Competing interests**

The authors declare no competing interests.

# Figure 1

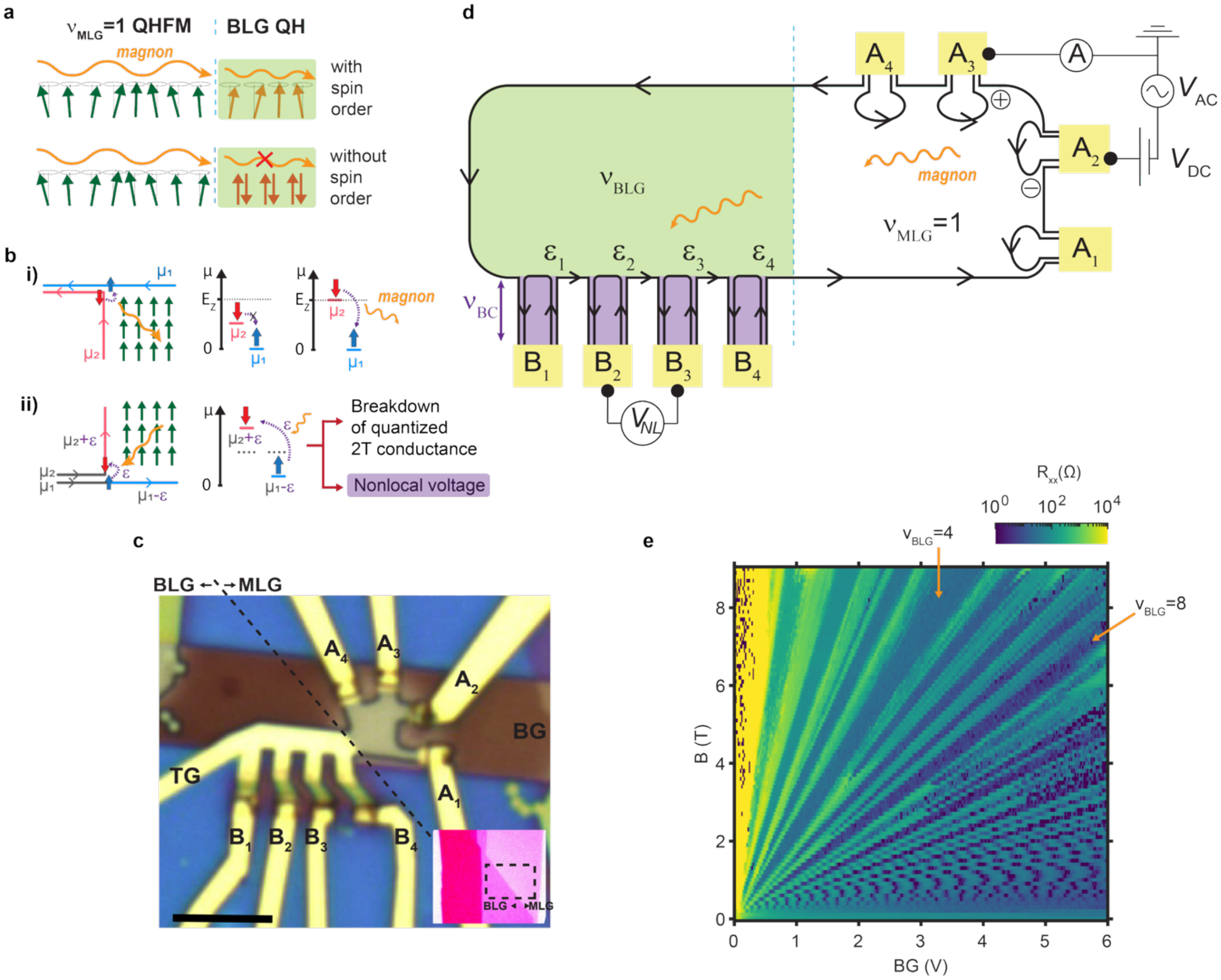

**Fig. 1 | Device schematics and characterization. a,** Illustration of magnon transmission across QHFM interfaces: magnons propagate across the MLG-BLG boundary only when magnetic order exists on both sides. **b,** Magnon generation (i) occurs when the chemical potential difference between oppositely spin-polarized edge states exceeds the Zeeman energy, enabling electron scattering from the hot edge to the cold edge and emitting magnons into the QHFM bulk. Magnon absorption (ii) induces electron scattering between spin-polarized edge states at the detection site, creating an imbalance in their chemical potentials that manifests as a breakdown of quantization in the two-terminal conductance and as a measurable nonlocal voltage at distant contacts. **c,** Optical micrograph of the hBN-encapsulated MLG/BLG device. A dashed line marks the MLG/BLG boundary, with contacts labeled A1-4 (MLG) and B1-4 (BLG). TG and BG denote top and bottom gates. Scale bar, 5 μm. Inset: Enhanced contrast image

showing MLG/BLG interface of the flake fabricated into the device. **d,** Experimental configuration for magnon transport measurements. Magnons are generated by applying DC bias $V_{DC}$ between two contacts in the MLG region. Generated magnons propagate through the MLG/BLG junction and the BLG bulk and are detected at the BLG contacts. In the main figures, contacts A2-A3 are used for magnon generation, and B2-B3 are used for magnon detection. **e,** Landau fan diagram showing longitudinal resistance between contacts B2 and B3 in the BLG region as a function of magnetic field B and back gate voltage, demonstrating well-developed integer and fractional quantum Hall states in BLG.

# Figure 2

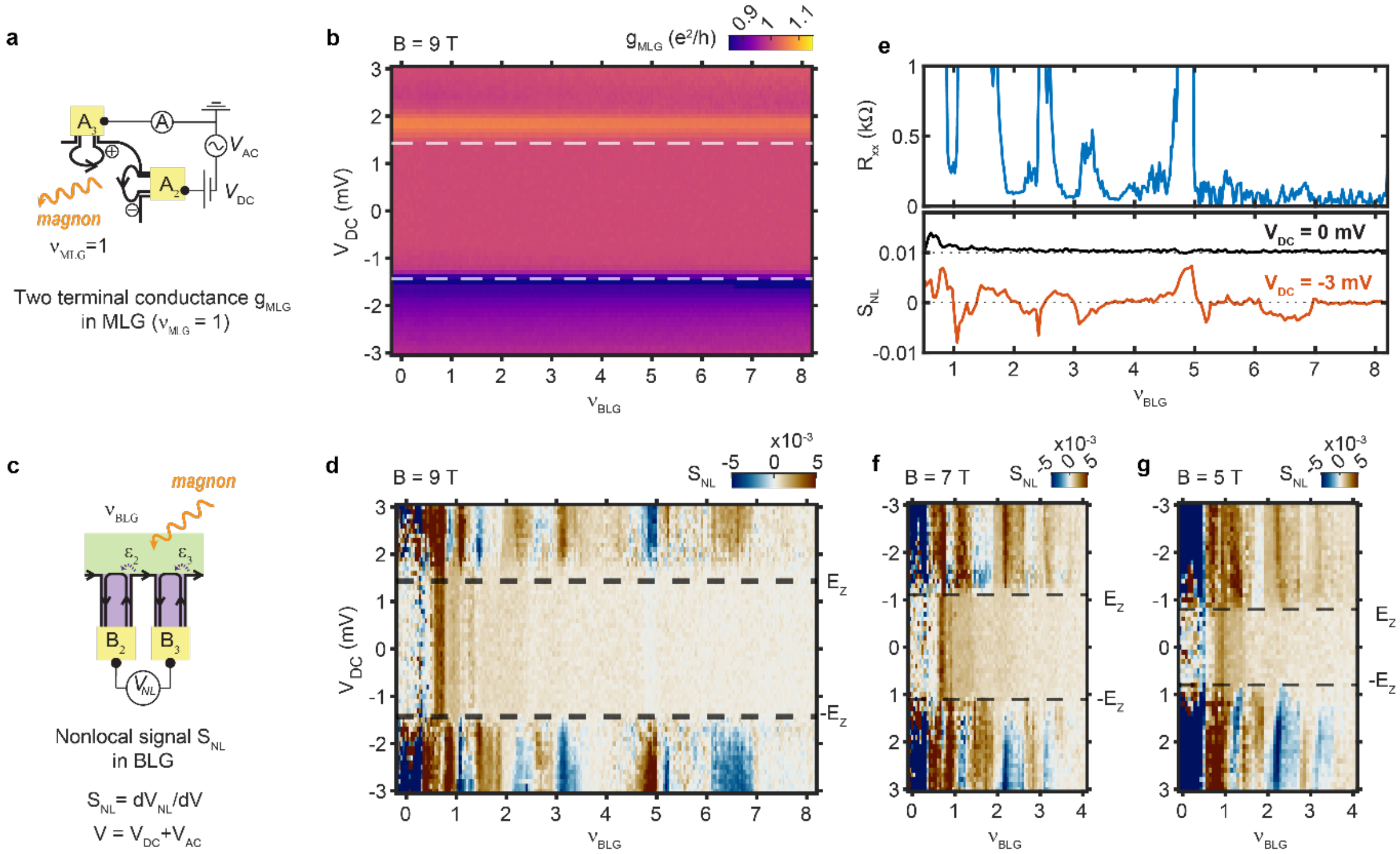

**Fig. 2 | Electrical detection of magnon transmission across MLG/BLG heterojunction. a,** Measurement schematic showing contacts for magnon generation and two-terminal conductance measurements in MLG. **b,** Two-terminal conductance $g_{MLG}$ in MLG region (contacts A2-A3) at $B$ = 9 T. MLG density is maintained within the $\nu_{MLG}$ = 1 plateau ($V_{BG}$ = 0.95 V) while $V_{DC}$ and $\nu_{BLG}$ are varied. Conductance quantization breaks down abruptly at the Zeeman energy $E_Z$ (white dashed lines), indicating that magnon generation starts when $|V_{DC}| > E_Z$ in MLG. **c,** Measurement schematic showing contacts for magnon detection with nonlocal voltage in BLG. **d,** Corresponding nonlocal signal $S_{NL} = dV_{NL}/dV$ in the BLG region (contacts B2-B3) at the same field $B$ = 9 T. Nonlocal signal appears sharply at the onset of $E_Z$ (black dashed lines), revealing that magnon transmission occurs across the MLG/BLG junction and BLG bulk. **e,** Direct comparison of longitudinal resistance (blue trace, measured between contacts B2-B3 at zero DC bias) with $S_{NL}$ (black and orange traces with $V_{DC}$ = 0 and -3 mV respectively. $V_{DC}$ = 0 mV trace is offset by 0.01 for visibility). The absence of correlation between the nonlocal signal and longitudinal resistance, combined with the disappearance of signal when $V_{DC}$ = 0 mV, confirms

the magnon-mediated origin of the nonlocal response. **f-g,** Nonlocal signal measured at $B$ = 7 T (**f**) and 5 T (**g**), demonstrating that the threshold consistently scales with the Zeeman energy across different field strengths.

# Figure 3

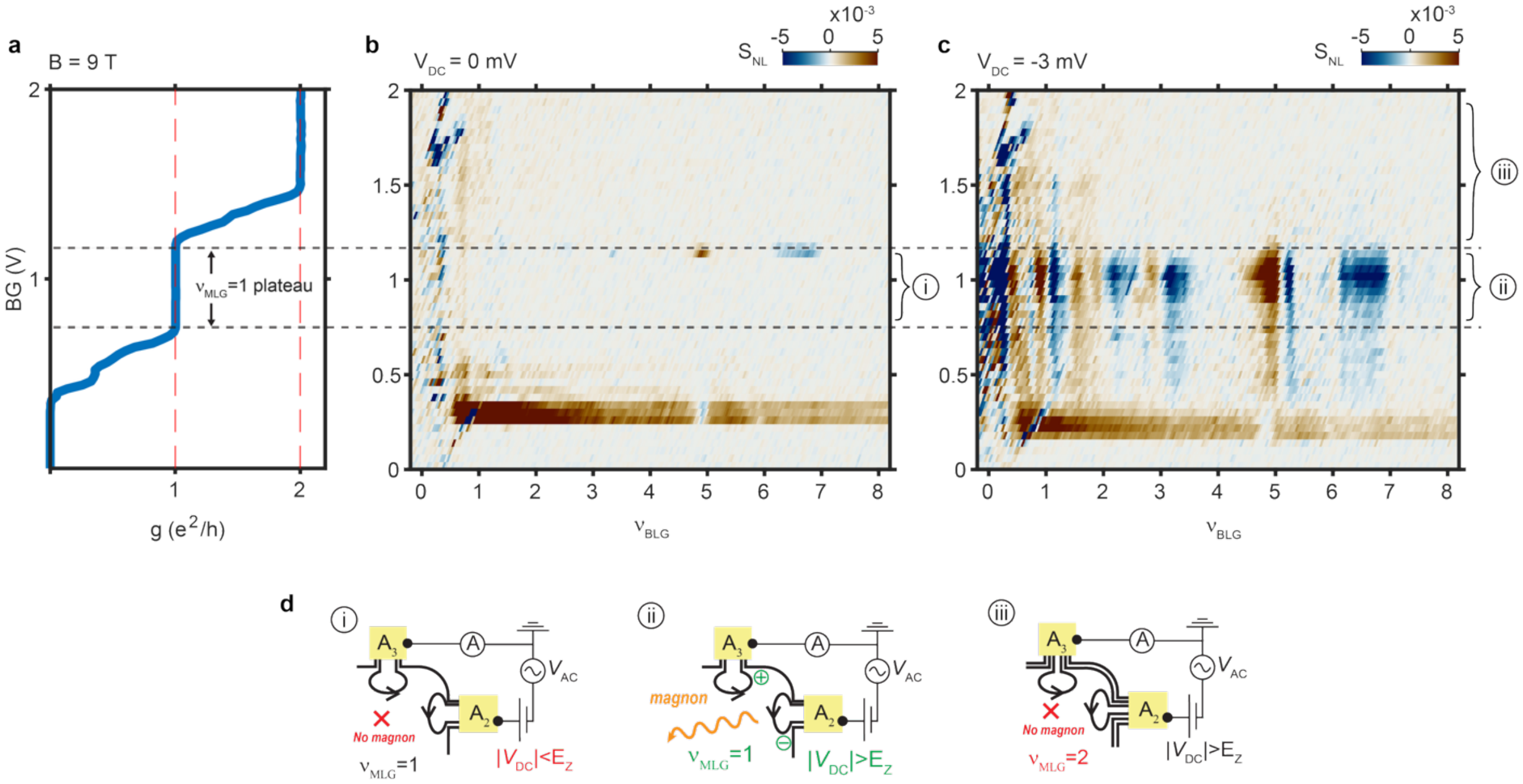

**Fig. 3 | Switching on/off magnon transmission signal with MLG density. a,** $g_{MLG}$ (contacts A2-A3) at $V_{DC}$ = 0 mV and $B$ = 9 T, showing a well-quantized plateau at $\nu_{MLG}$ = 1 and 2. **b-c,** Nonlocal signal $S_{NL}$ in the BLG at $B$ = 9 T with $V_{DC}$ = 0 mV (magnon generation off) (**b**) and $V_{DC}$ = -3 mV (magnon generation on) (**c**). The measurement configuration is identical to Fig. 2 with A1 and A4 grounded. Nonlocal signal is absent when $V_{DC}$ = 0 mV, but it emerges robustly when $V_{DC}$ = -3 mV. The signal is strongest precisely when BG value is within the $\nu_{MLG}$ = 1 plateau region, where magnon generation is most prominent. The signal vanishes as the MLG density is tuned away from $\nu_{MLG}$ = 1 plateau toward $\nu_{MLG}$ = 2 plateau, where the system becomes nonmagnetic, resulting in suppression of magnon generation and propagation in MLG. **d,** Schematics illustrating three regimes presented in panels a-c: (i) No magnon generation at $\nu_{MLG}$ = 1 when $|V_{DC}| < E_Z$. (ii) Magnon generation at $\nu_{MLG}$ = 1 when $|V_{DC}| > E_Z$, where magnons propagate from MLG into BLG. (iii) No magnon generation at $\nu_{MLG}$ = 2 even when $|V_{DC}| > E_Z$, as the nonmagnetic bulk suppresses magnon generation and propagation.

# Figure 4

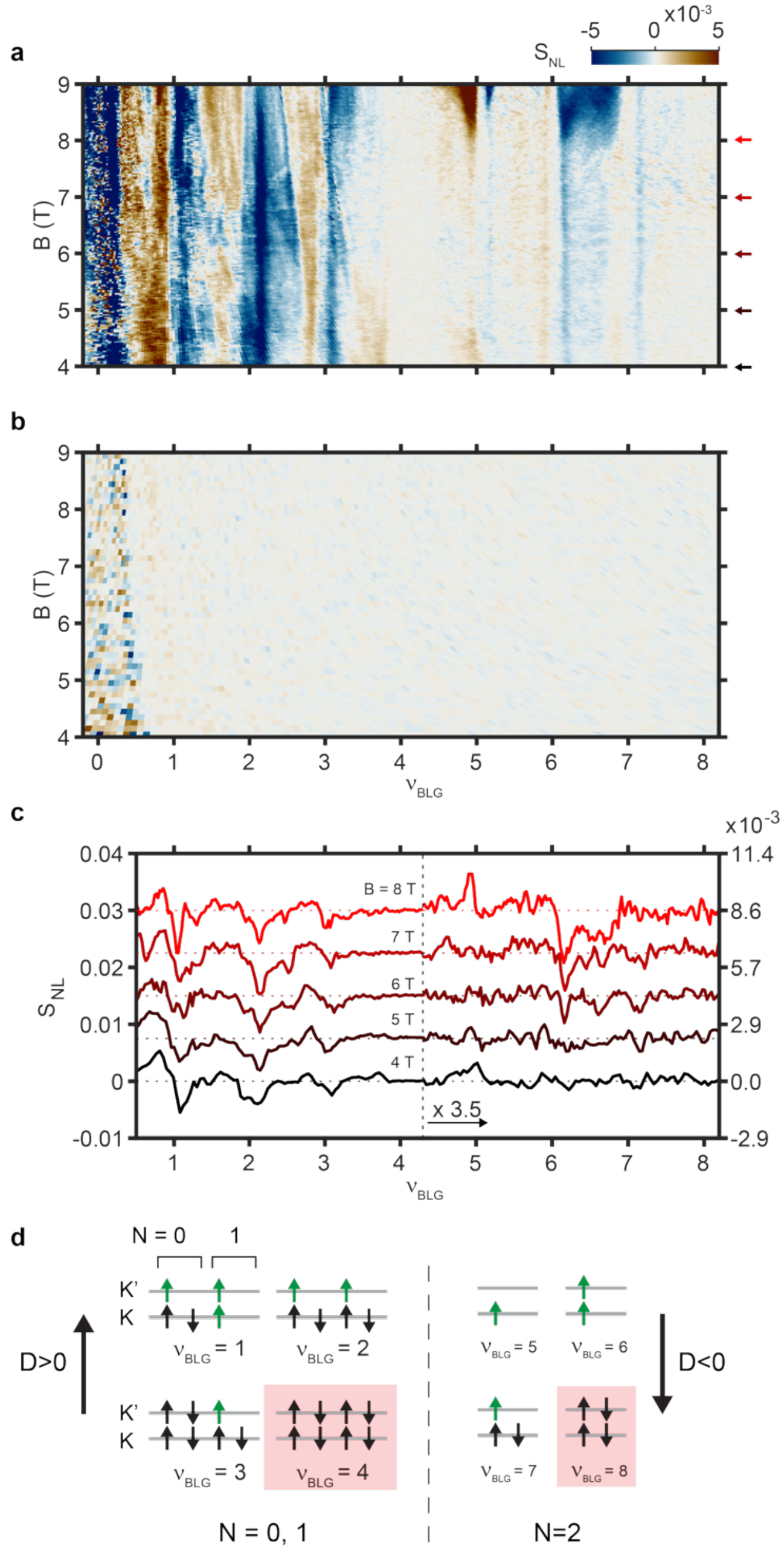

**Fig. 4 | Magnetic field dependence of magnon transmission in BLG quantum Hall states. a-b,** Nonlocal signal $S_{NL}$ measured as magnetic field $B$ is varied from $B = 4$ T to 9 T for positive fillings $\nu_{BLG}$ at **a,** $V_{DC} = -3$ mV (generating magnon) and **b,** $V_{DC} = 0$ mV (no magnons). Configuration is the same as in the Fig. 2. Magnon transmission is absent at $V_{DC} = 0$ mV. When

$|V_{DC}| > E_Z$, the nonlocal signal is clearly observed to exhibit absorption features at constant filling factors, appearing as vertical lines. Nonlocal signals are observed at the broken-symmetry fillings $0 < \nu_{BLG} < 4$ and for $4 < \nu_{BLG} < 8$, indicating these states specifically break spin symmetry and therefore support magnon transmission. The transmission disappears at the full fillings ($\nu_{BLG} = 4, 8$) where the spin symmetry is recovered and magnons thus cannot propagate. **c,** Linecuts of the nonlocal signal at several values of $B$. Each line is vertically offset for visibility, with dotted lines marking where $S_{NL} = 0$. The magnitude of $S_{NL}$ is diminished in the $N = 2$ LL, and the trace is therefore multiplied by a factor of 3.5 to enhance visibility for $\nu_{BLG} > 4.3$. The field value of each linecut is indicated by the arrow of the same color on the right side of panel a. **d,** Schematic isospin ordering at different filling factors, showing valley (K/K'), spin (↑↓), and orbital ($N = 0, 1, 2$) polarization. $D$ shows the displacement direction ($D > 0$ for $N = 0,1$ and $D < 0$ for $N = 2$) matching experimental condition. This ordering is consistent with magnon transmission at broken-symmetry fillings $0 < \nu_{BLG} < 4$ and $4 < \nu_{BLG} < 8$ and its suppression at the symmetric fillings $\nu_{BLG} = 4, 8$.

# Extended Data Figure 1

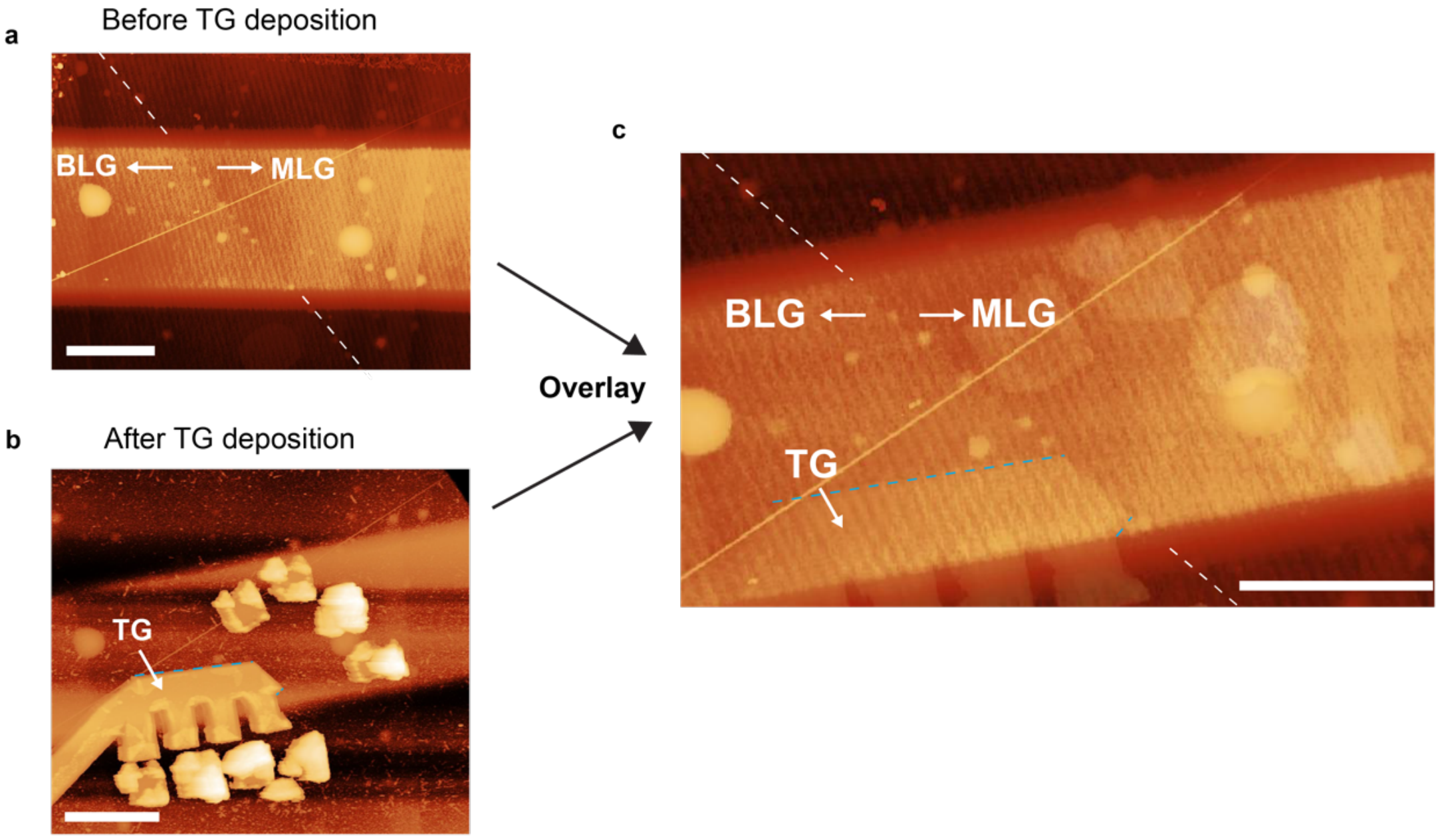

**Extended Data Fig. 1 | AFM images of MLG/BLG boundary alignment with top gate (TG).** Scale bars: 3 µm. **a,** AFM image of the hBN-encapsulated MLG/BLG stack for device A before TG evaporation. The stack is placed above the PdAu metal gate (horizontally extended rectangular feature in the middle of the image). White dashed lines are drawn along the MLG/BLG boundary as a guide to the eye. **b,** AFM image after TG and contact evaporation. Sky-blue dashed lines mark the TG edges that connect to the vertices of the edge aligned to the MLG/BLG boundary. **c,** Overlaying the two images confirms precise alignment between the TG edge and the MLG/BLG boundary, which enables independent carrier density control of the MLG and BLG regions.

# Extended Data Figure 2

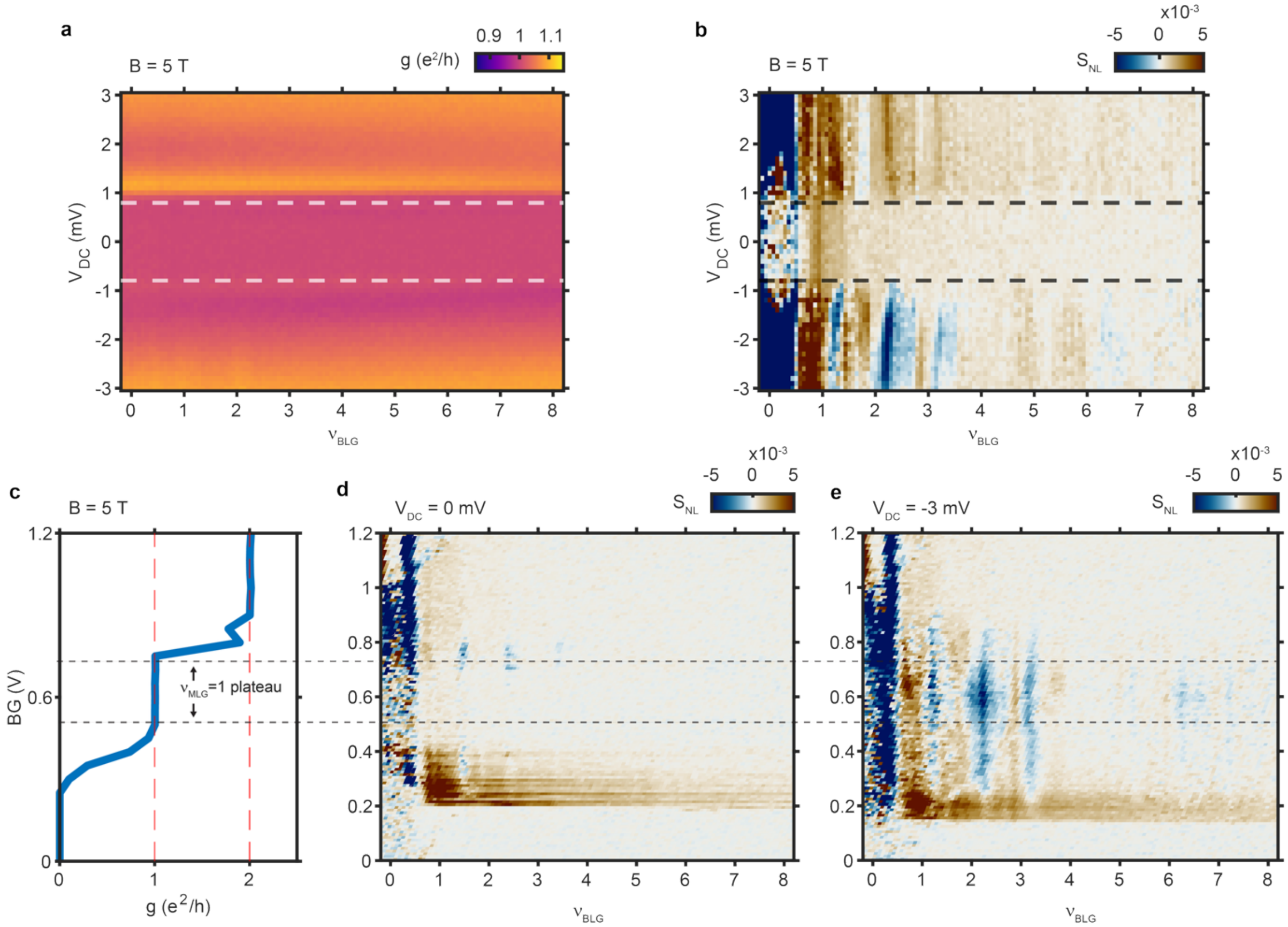

**Extended Data Fig. 2 | Electrical detection of magnon transmission across MLG/BLG heterojunction at B = 5 T. a,** Two-terminal conductance $g_{MLG}$ in MLG region (contacts A2-A3). MLG density is set inside the $\nu_{MLG} = 1$ plateau ($V_{BG} = 0.613$ V) and $V_{DC}$ and $\nu_{BLG}$ were varied. The conductance breakdown occurs at $E_Z$ (white dashed lines) corresponding to $B = 5$ T. **b,** Corresponding nonlocal signal $S_{NL}$ in BLG region (contacts B2-B3). As in Fig. 2, the nonlocal signal appears sharply at $E_Z$ (black dashed lines), confirming that the phenomenology properly scales with Zeeman energy. The observation of this consistent scaling of both conductance breakdown and nonlocal signal with Zeeman energy across different magnetic fields provides strong evidence that the phenomenon originates from magnon generation and transmission. **c-e,** The dependence of the magnon transmission signal on the MLG carrier density. **c,** MLG two-terminal conductance at $V_{DC} = 0$ mV. **d-e,** Nonlocal signal in BLG at $V_{DC} = 0$ mV (magnon generation off) (**d**) and $V_{DC} = -3$ mV (magnon generation on) (**e**). A1 and A4 are grounded. The

nonlocal signal is absent when $V_{DC}$ = 0 mV but clearly appears when $V_{DC}$ = -3 mV. The data reproduce the key features observed at 9 T (Fig. 3): absence of the signal at $V_{DC}$ = 0 mV, clear transmission when $|V_{DC}| > E_Z$, and confinement of signal to the $\nu_{MLG}$ = 1 plateau region where magnon generation and propagation are most prominent.

# Extended Data Figure 3

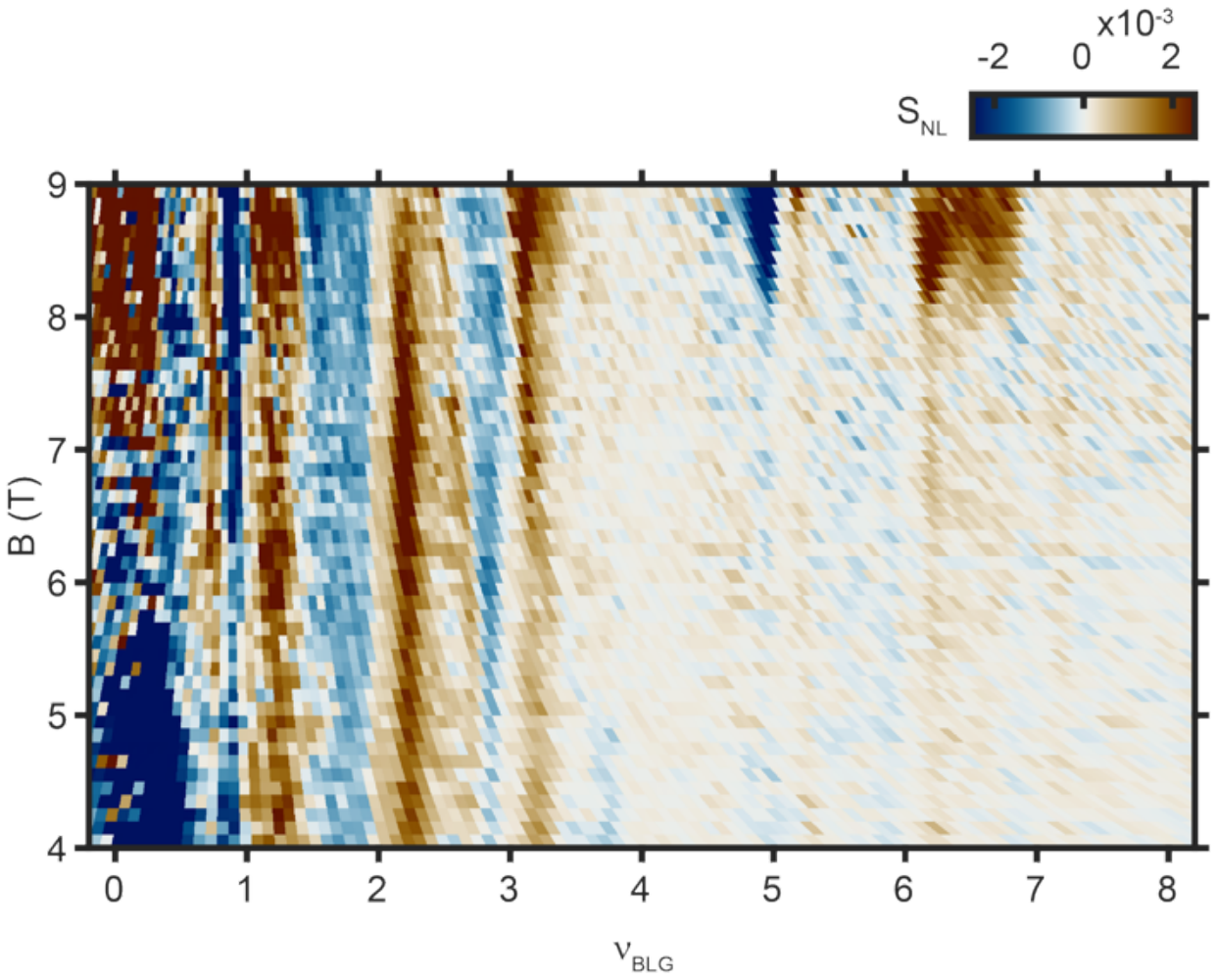

**Extended Data Fig. 3 | Magnetic field dependence of the magnon transmission in BLG quantum Hall states with $V_{DC}$ = +3 mV.** Nonlocal signal measured as a function of magnetic field $B$ (4 T to 9 T) and BLG filling factor $\nu_{BLG}$ with a DC bias $V_{DC}$ = +3 mV applied to the magnon generation contacts. Contacts A1 and A4 are grounded. The observed phenomenology reproduces the behavior when $V_{DC}$ = -3 mV was applied (Fig 4a), demonstrating the robustness of the magnon transmission signal independent of the particular value of $V_{DC}$ chosen.

# Extended Data Figure 4

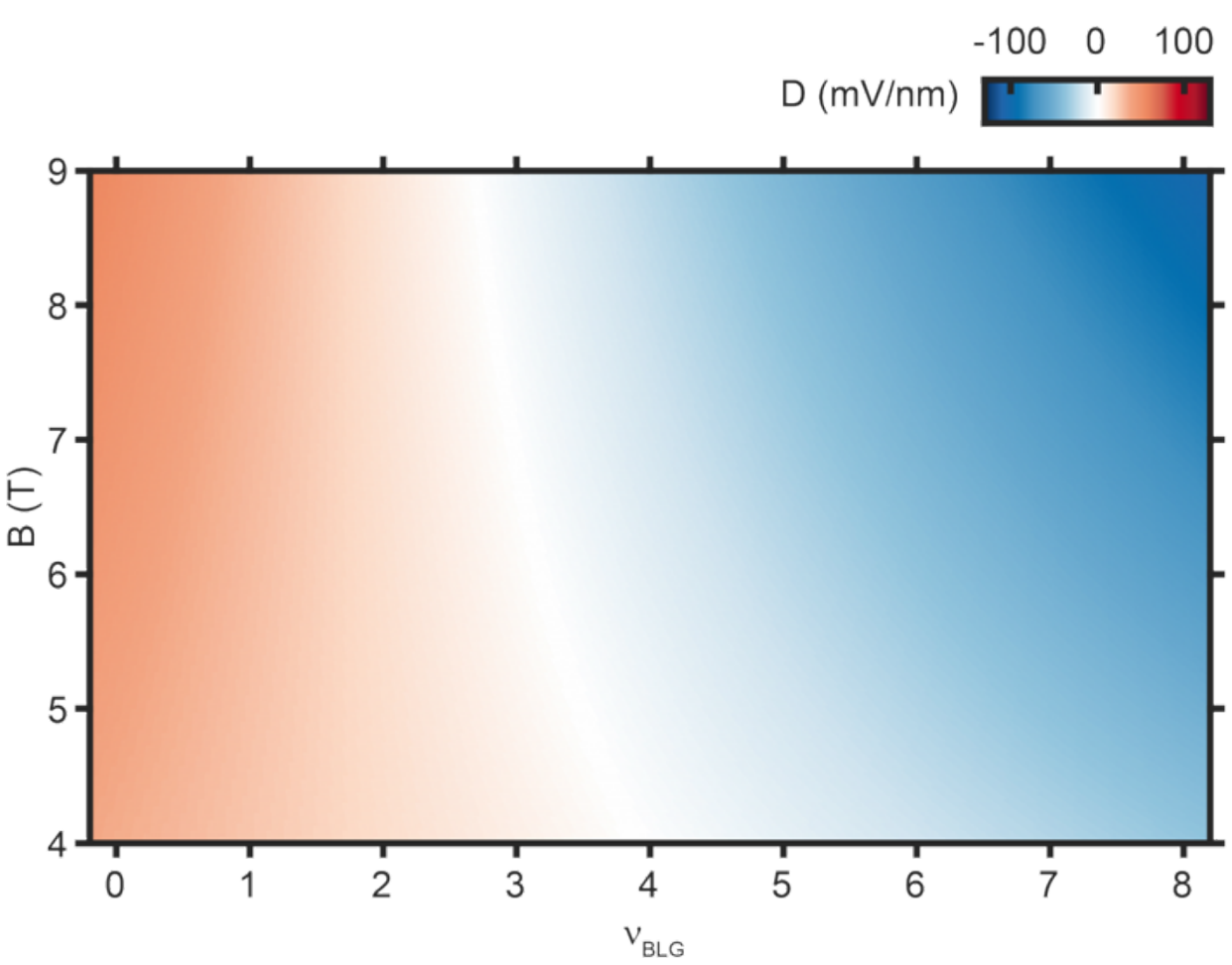

**Extended Data Fig. 4 | Displacement field in BLG during magnetic field sweep measurements.** Displacement field *D*, calculated from extracted top and bottom gate capacitances during the magnetic field scan presented in Fig. 4. $|D|$ remains below 100 mV/nm for all presented field sweep measurements.

# Extended Data Figure 5

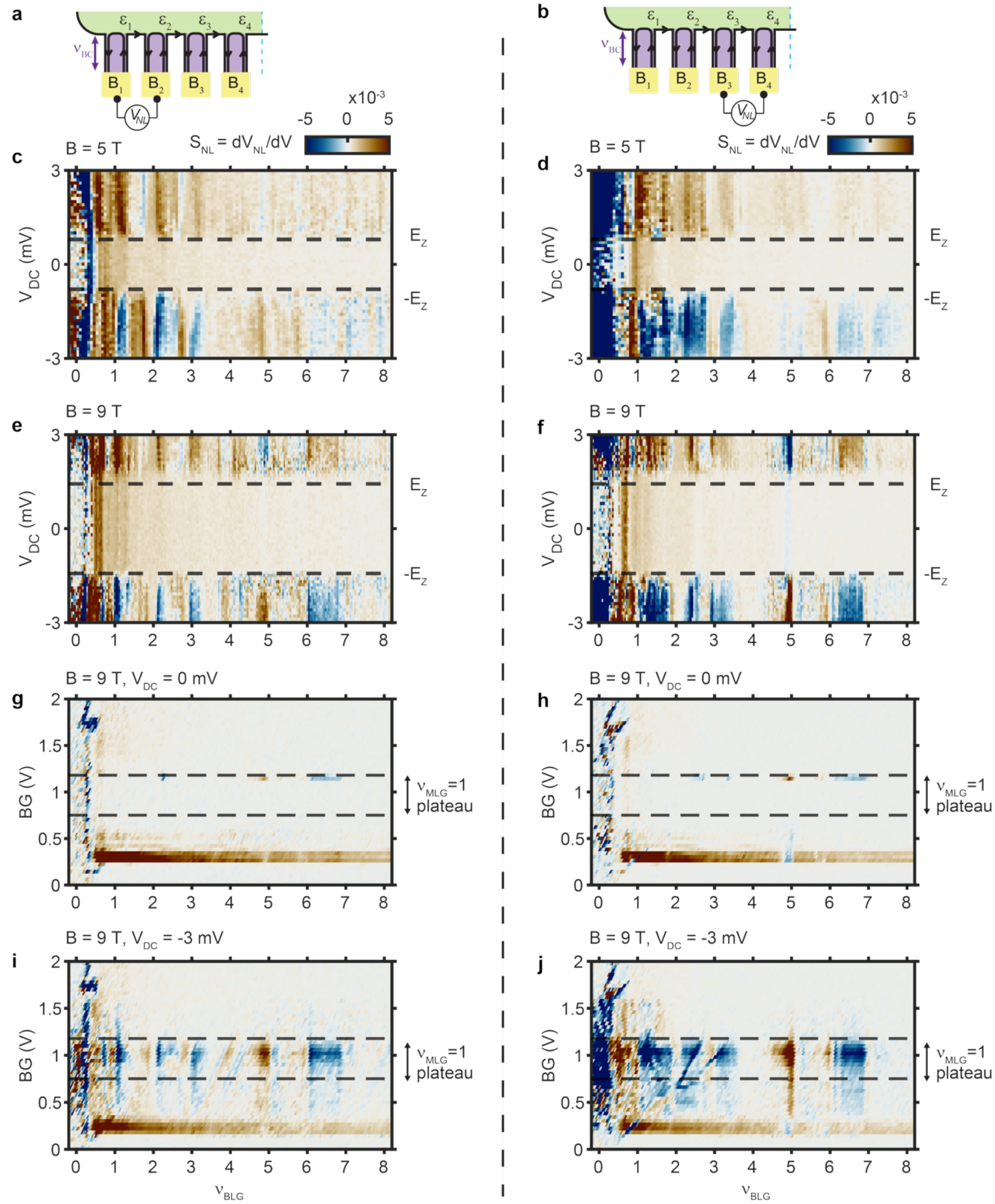

**Extended Data Fig. 5 | Magnon transmission across MLG/BLG junction probed with different BLG contact pairs. a-b,** Schematics showing pairs of contacts used for the corresponding data. The contacts indicated in panel a (B1, B2) are used in the measurements shown in the left column (c, e, g, i), the contacts indicated in panel b (B3, B4) are used in the

measurements shown in the right column (d, f, h, j). **c-f** Nonlocal signals measured at $B$ = 5 T (c-d) and 9 T (e-f), with MLG density set inside $\nu_{MLG}$ = 1 plateau ($V_{BG}$ = 0.613 V at 5 T and $V_{BG}$ = 0.95 V at 9 T). For both values of $B$ and for both pairs of contacts, the nonlocal signal appears abruptly at the Zeeman energy $E_Z$ (white dashed lines). **g-j,** Nonlocal signal as a function of $\nu_{BLG}$ and $V_{BG}$ at $B$ = 9 T with $V_{DC}$ = 0 mV (g, h) and $V_{DC}$ = -3 mV (i, j). Black dashed lines indicate the $V_{BG}$ range where MLG two-terminal conductance $g_{MLG}$ (not shown) remains at the $\nu_{MLG}$ = 1 plateau.

# Extended Data Figure 6

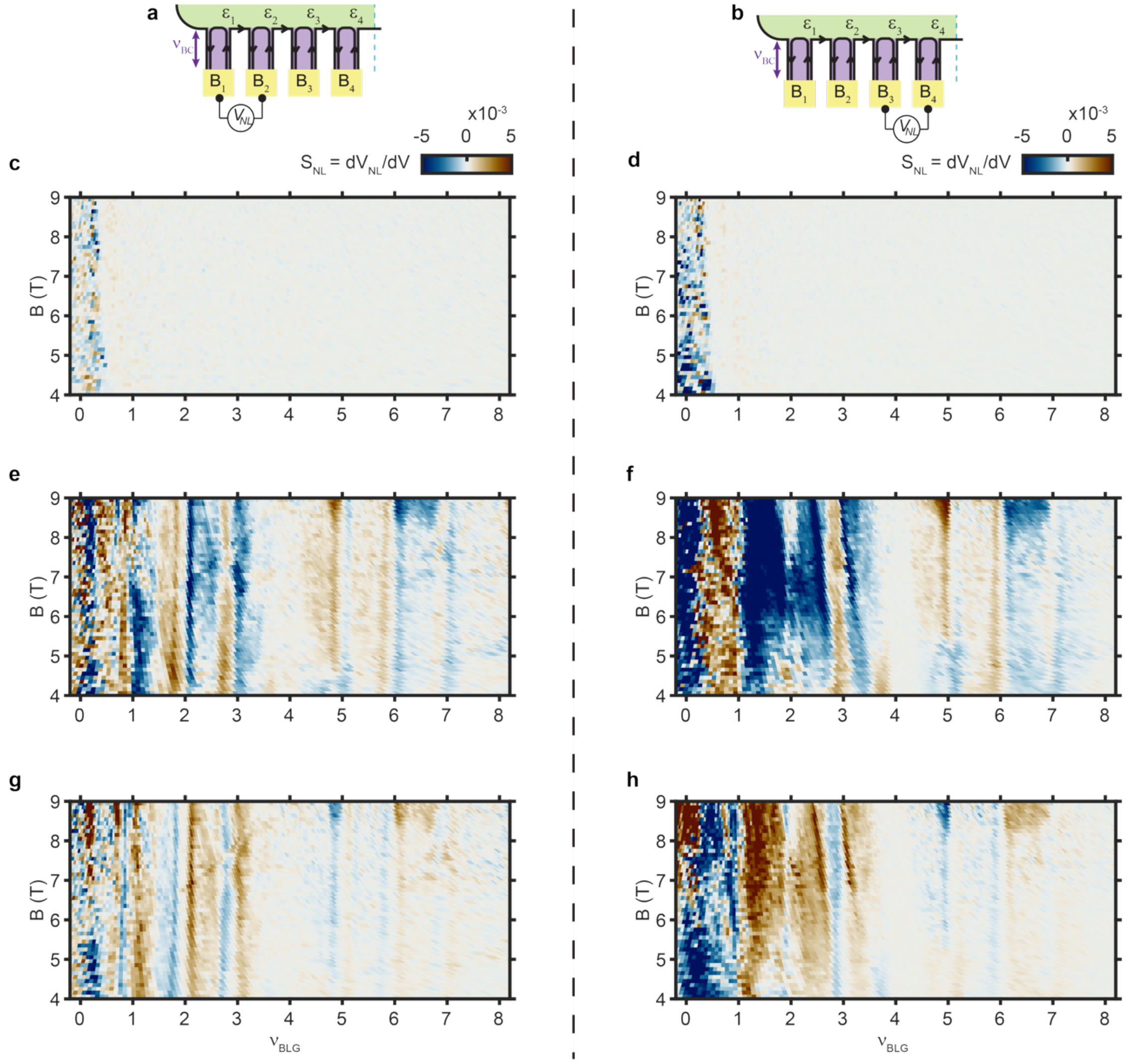

**Extended Data Fig. 6 | Magnetic field dependence of magnon transmission probed with different BLG contact pairs. a-b,** Schematics showing pairs of contacts used for the corresponding data. The contacts indicated in panel a (B1, B2) are used for the measurements in the left column (c, e, g), while the contacts indicated in panel b (B3, B4) are used for the measurements in the right column (d, f, h). **c-h,** Nonlocal signal measured as magnetic field $B$ is varied from 4 T to 9 T for positive fillings $\nu_{BLG}$ when $V_{DC}$ = 0 mV (c, d), $V_{DC}$ = -3 mV (e, f), and $V_{DC}$ = +3 mV (g, h). Consistently, the measurements show the absence of signal without magnon

generation and robust magnon transmission with DC bias above the Zeeman threshold. The key features observed in Fig. 4 are reproduced with both contact pairs: the nonlocal signal follows Landau level filling, appearing as vertical lines in $B$ vs $\nu_{BLG}$ parameter space, is present at symmetry broken fillings in $0 < \nu_{BLG} < 8$, and disappears at full fillings ($\nu_{BLG} = 4, 8$). This consistency across different contact configurations is consistent with the bulk nature of magnon transmission through BLG QHFM states.

# Extended Data Figure 7

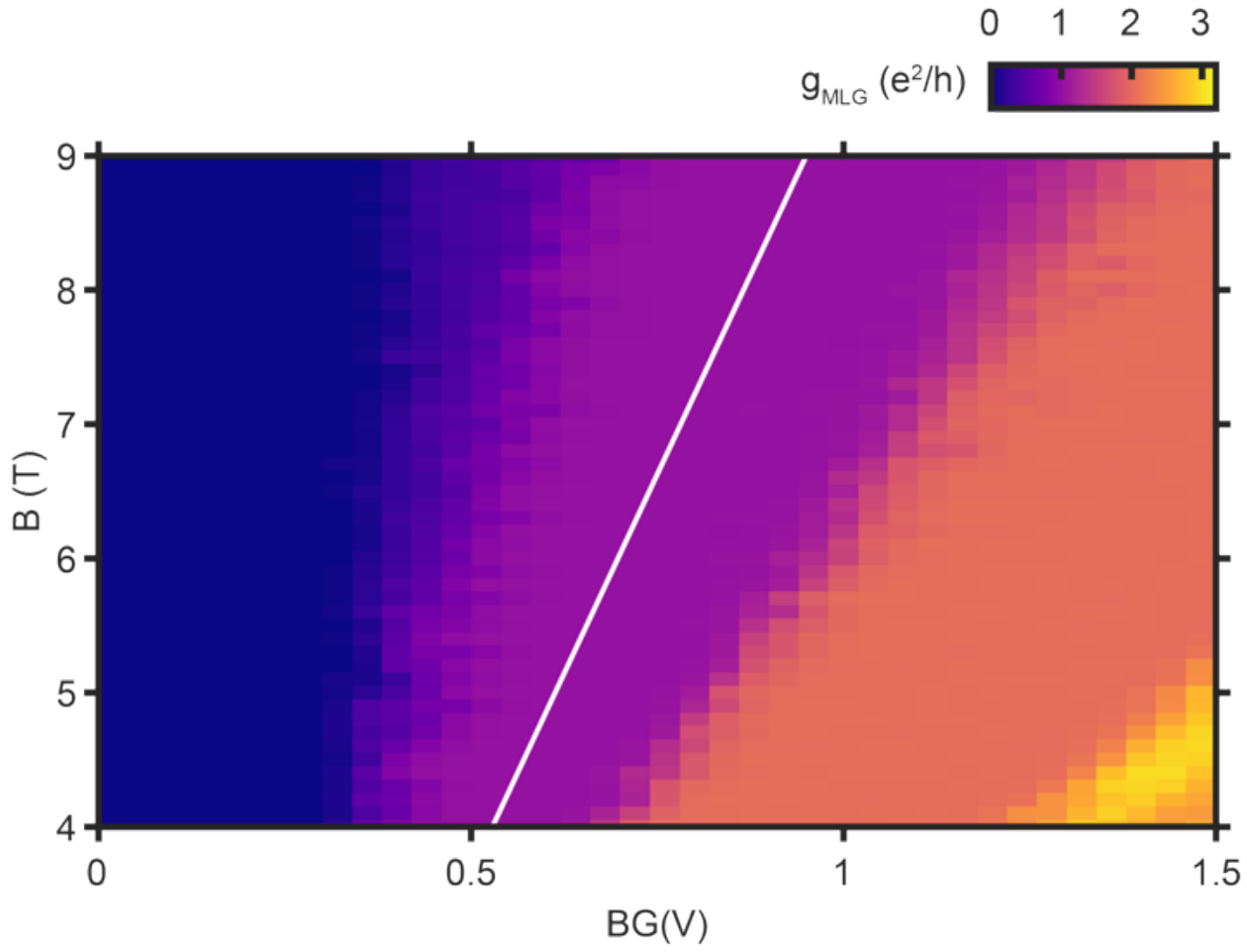

**Extended Data Fig. 7 | Maintenance of $\nu_{MLG} = 1$ filling during magnetic field-dependent measurements.** Two-terminal conductance $g_{MLG}$ plotted as a function of back-gate voltage $V_{BG}$ and magnetic field $B$. The white line traces the back-gate voltage values used for the nonlocal measurements in Fig. 4, demonstrating that the MLG region remains within the $\nu_{MLG} = 1$ quantum Hall plateau throughout the field sweep. This ensures consistent magnon generation conditions across the entire range of magnetic fields studied.

# Extended Data Figure 8

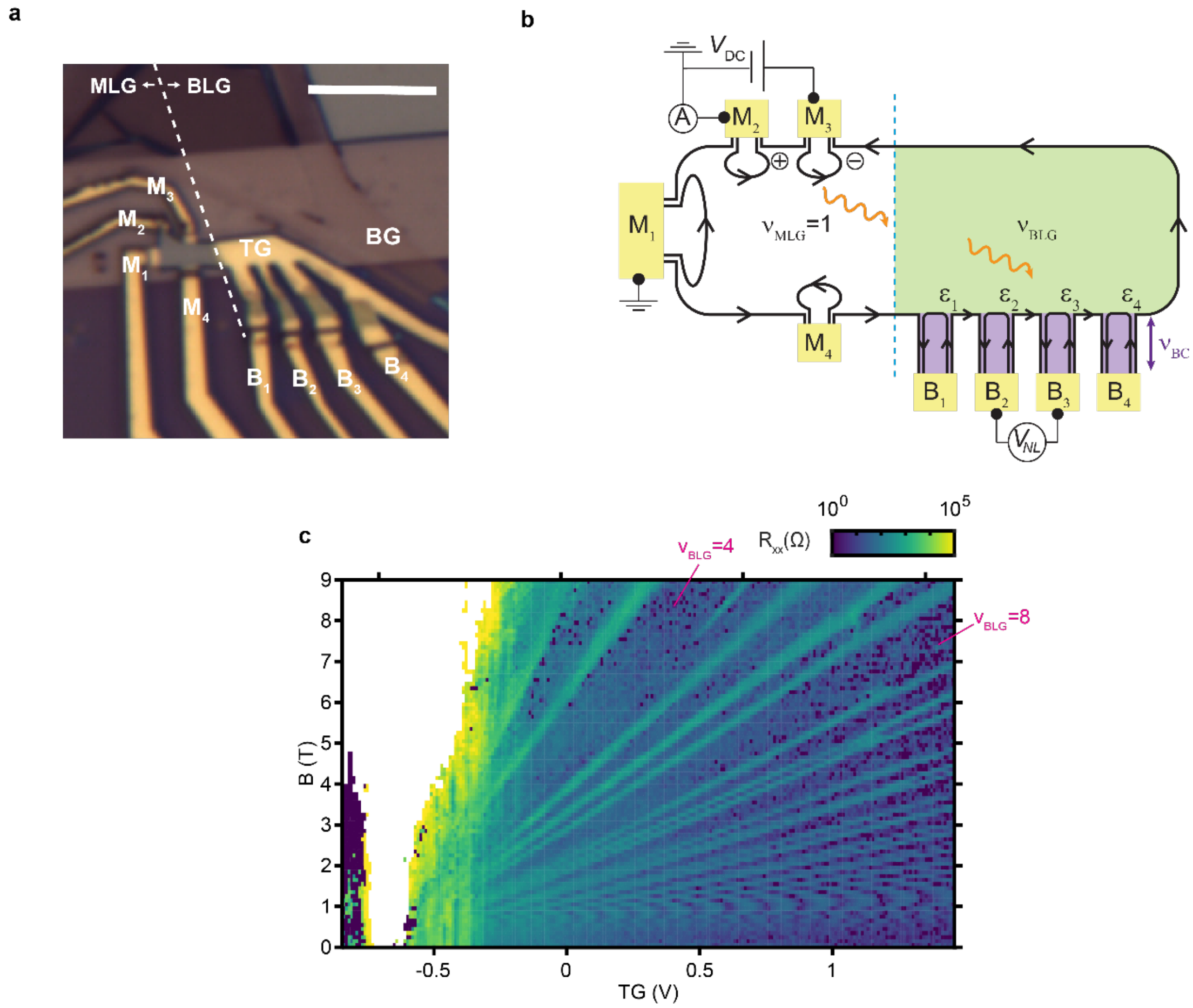

**Extended Data Fig. 8 | Device B schematic and characterization. a,** Optical micrograph of device B. A dashed line marks the MLG/BLG boundary, with the MLG and BLG regions labeled. TG and BG denote top and bottom gates. Scale bar, 5 μm. **b,** Experimental configuration for magnon transport measurements. Magnons are generated by applying a DC bias $V_{DC}$ between contacts M2 and M3 in the MLG region, propagate across the MLG/BLG boundary and the BLG bulk, and are detected at the BLG contacts B2 and B3. **c,** Landau fan diagram showing longitudinal resistance in the BLG sector with magnetic field $B$ versus top gate (TG) voltage with the back gate held at 2 V. Well-developed integer quantum Hall states demonstrate the high quality of the device.

# Extended Data Figure 9

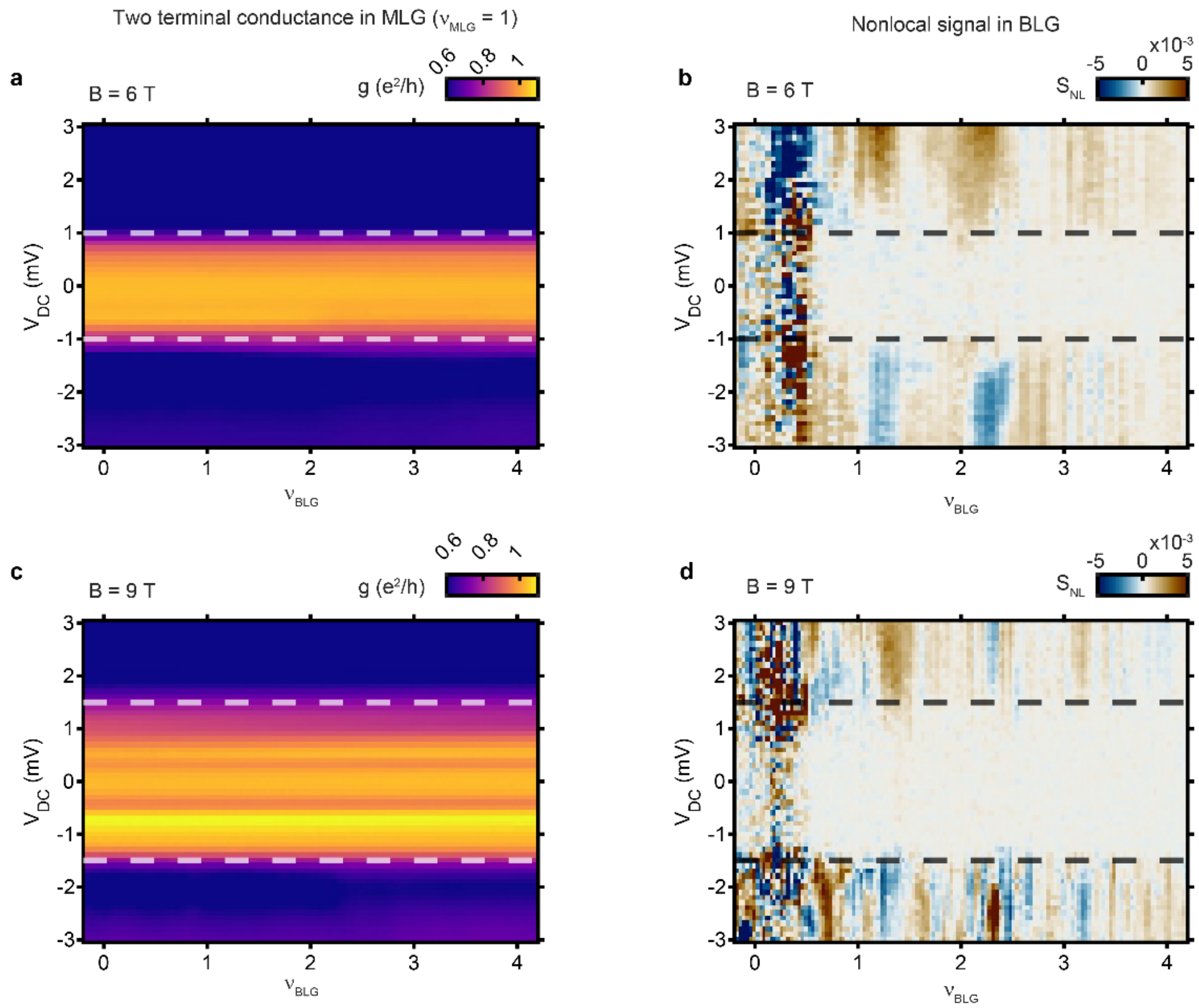

**Extended Data Fig. 9 | Zeeman onset of magnon transmission in Device B at different magnetic fields. a-b,** Measurements at $B = 6$ T. The measurement configuration is the same as that shown in Extended Data Fig 8. **a,** Two-terminal conductance $g_{MLG}$ in MLG tuned to $\nu_{MLG} = 1$ plateau, showing breakdown of the quantized conductance as a function of DC bias $V_{DC}$ and BLG filling factor $\nu_{BLG}$. The onset occurs at the Zeeman energy $E_Z$ (white dashed lines). **b,** Corresponding nonlocal signal $S_{NL}$ in BLG region demonstrates sharp onset at $E_Z$ (black dashed lines), indicating that magnon transmission occurs across MLG/BLG junction and BLG bulk. **c-d,** Identical measurements at $B = 9$ T showing the onset follows Zeeman energy, as expected for Zeeman-activated magnon generation.

# Extended Data Figure 10

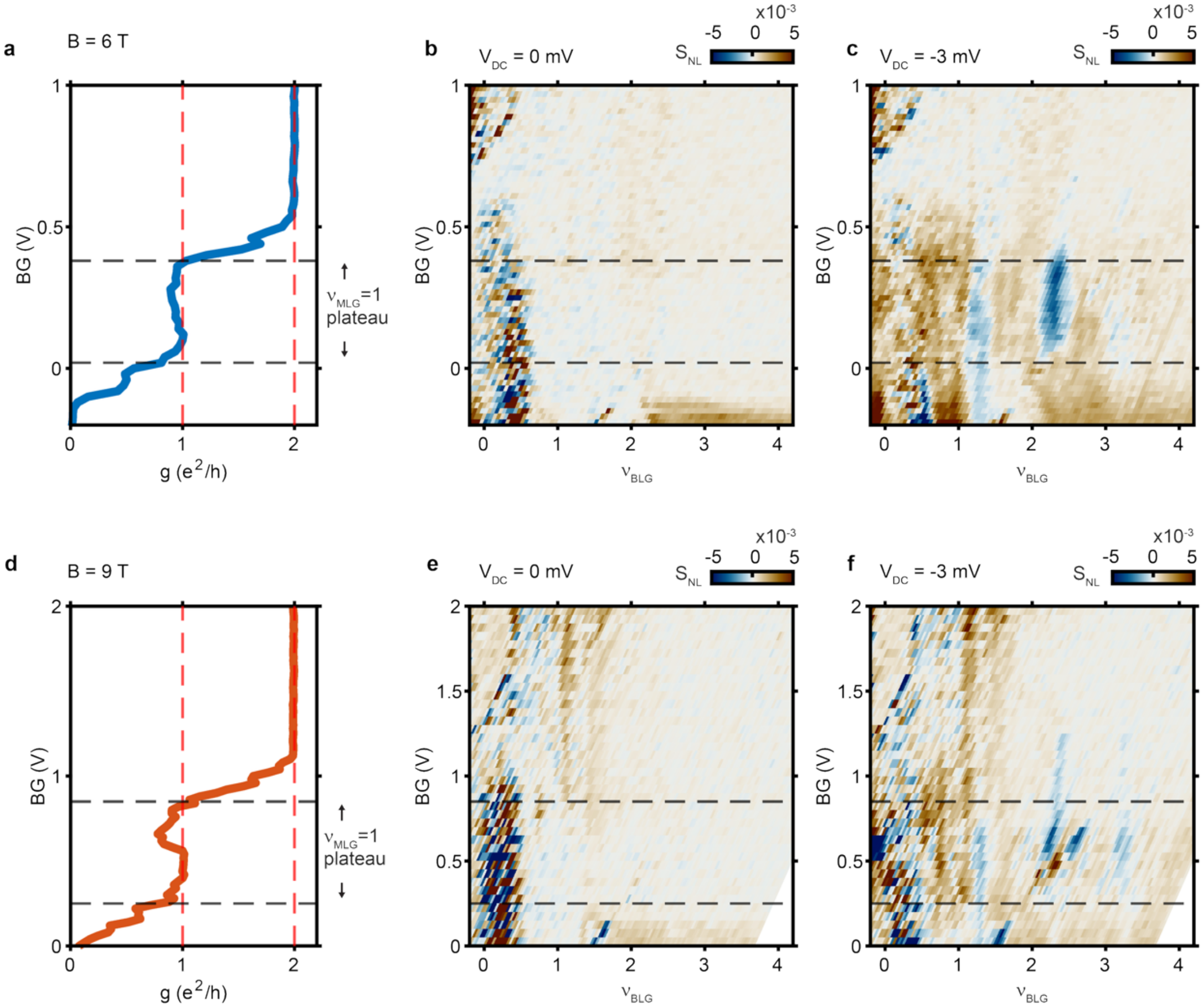

**Extended Data Fig. 10 | MLG carrier density dependence of magnon transmission at $B$ = 6 T and 9 T in Device B. a-c,** Dependence of magnon transmission signal on MLG density at $B$ = 6 T. The measurement configuration is the same as that shown in Extended Data Fig 8. **a,** MLG two-terminal conductance at $V_{DC}$ = 0 mV. **b, c,** Nonlocal signal in BLG at (**b**) $V_{DC}$ = 0 mV (magnon generation off) and (**c**) $V_{DC}$ = -3 mV (magnon generation on). Nonlocal signal is absent at $V_{DC}$ = 0 mV but emerges clearly when $V_{DC}$ = -3 mV, with the signal most enhanced within the $\nu_{MLG}$ = 1 plateau region (black dashed lines) where magnon generation is most prominent. The signal vanishes as MLG density approaches $\nu_{MLG}$ = 2, consistent with suppression of magnon generation and propagation. **d-f,** The same set of measurements demonstrating consistent behavior at higher magnetic field $B$ = 9 T.

# Extended Data Figure 11

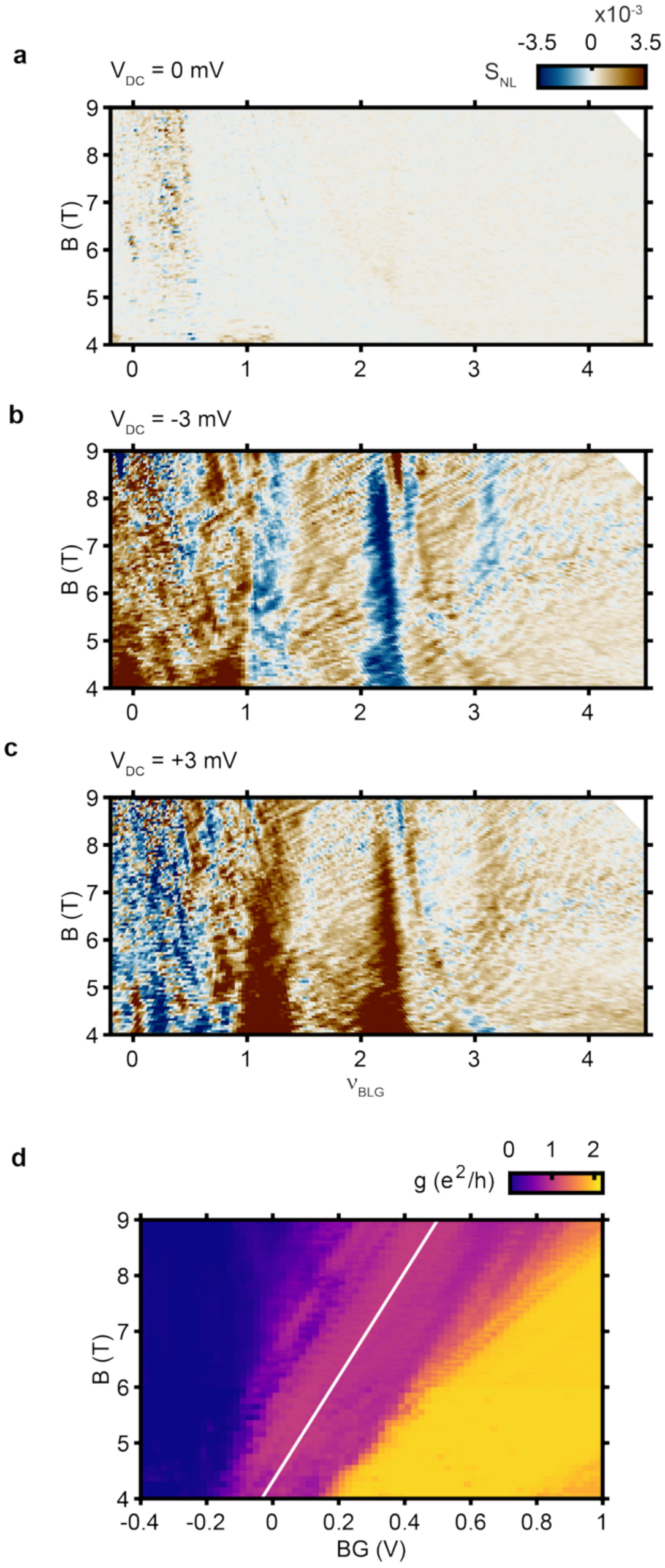

**Extended Data Fig. 11 | Magnetic field dependence of magnon transmission in Device B.** Measurement configuration is the same as that shown in Extended Data Fig. 8. **a-c,** Nonlocal signal measured as magnetic field $B$ is varied for positive fillings $\nu_{BLG}$ at **a,** $V_{DC} = 0$ mV (no magnons), **b,** $V_{DC} = -3$ mV, and **c,** $V_{DC} = +3$ mV. The nonlocal signal is absent when $V_{DC} = 0$ mV but emerges when $|V_{DC}| > E_Z$. The signal exhibits prominent vertical lines, indicating

features associated with magnon absorption occurring at fixed Landau level filling factor. Nonlocal signals are observed at symmetry broken fillings in $0 < \nu_{BLG} < 4$ which support magnon transmission, while disappearing at full filling ($\nu_{BLG} = 4$) where spin degrees of freedom freeze out. **d,** MLG two-terminal conductance plotted as a function of back gate voltage BG and magnetic field *B*. The white line traces the back gate voltage values used during nonlocal measurements in panels a-c, demonstrating that the MLG region remains within the $\nu_{MLG} = 1$ quantum Hall plateau throughout the field sweep to ensure consistent magnon generation conditions.